# Novel design of refractive index sensors and bio-sensors based on a dual-core micro-structured optical fiber

G. N. Tsigaridas[1,*], V. Karvouniaris[2], G. Chalkiadakis[1] and P. Persephonis[2]

[1] *Department of Physics, School of Applied Mathematical and Physical Sciences, National Technical University of Athens, Zografou Campus, GR-15780 Zografou, Athens, Greece*

[2] *Department of Physics, University of Patras, GR-26504 Rion Patras, Greece*

[*]Corresponding author: G. N. Tsigaridas     E-mail: gtsig@mail.ntua.gr

**Abstract:** In the present work a new model of a refractive index (RI) sensor is exhibited. This is based on a dual core micro-structured optical fiber (MOF), where two holes are introduced at the core centers. In this way, the model enhances the interaction of the fiber modes propagated in the core region, providing the possibility of increasing the dimensions of the fiber sensor. Thus, the filling of the fiber holes with the fluid under study is facilitated, and generally the practical use of the system as a refractive index sensor is greatly simplified. The sensitivity of the system for various configurations has also been determined. It is found that it can reach record values of the order of 7000 nm/RIU. Finally, the use of the system as a bio-sensor has been examined, giving very promising results regarding both the sensitivity and the ease of the calibration.

## 1. Introduction

Micro-structured optical fibers (MOF) are very well suited for refractive index (RI) sensing [1-12] and bio-sensing [13-20] applications,. In most cases, they consist of an array of holes running along the entire length of the fiber, surrounding a solid or hollow core. In the first case the guiding is achieved by photonic bandgap, while in the second one by total internal reflection. A special class of micro-structured optical fibers is the dual solid core ones [3, 18]. In this case, coupling between the propagating modes in the two cores occurs, because of their interaction through their evanescent fields. This coupling can also be described using a pair of super-modes, (a symmetric (even) and an anti-symmetric (odd) one), which are depicted in Fig. 1.

In all cases, the RI sensing is attained by filling the fiber holes with the fluid under investigation and observing the changes in the propagating modes. However, from a practical point of view the filling of the fiber holes implies difficulties because of their small diameter, typically of the order of $1\ \mu m$. Therefore, it would be desirable to increase the dimensions of the fiber holes in order to facilitate the practical application of the sensor. However, this implies that the distance between the fibers cores should also increase, decreasing the strength of their interaction. In the present work, a modification of the basic design is proposed, by introducing a hole at the center of the fiber cores. This results to spreading of the mode field and increases the interaction between the cores, providing the opportunity to increase their separation without suppressing their interaction.

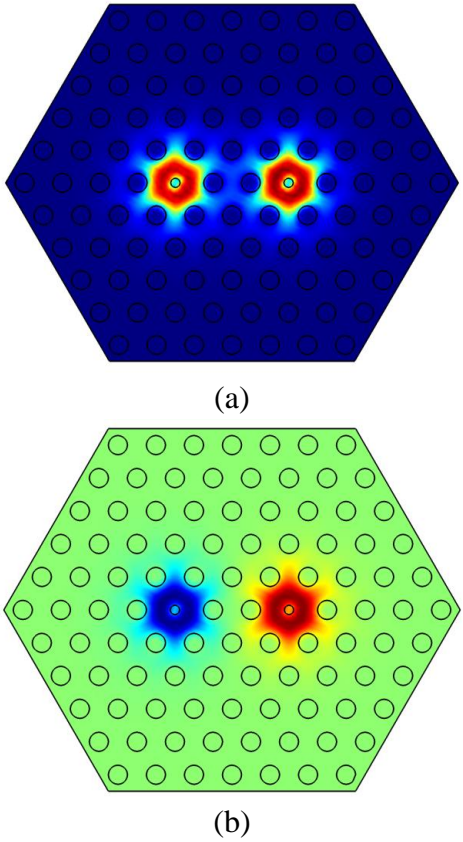

**Fig. 1**: The electric field distribution for a symmetric (a) and an anti-symmetric (b) fiber mode.

From a physical point of view, the coupling is due to a periodic transfer of power between the two cores and can be expressed by the formula [3, 18]

$$T_{21} = \sin^2\left(\frac{\theta}{2}\right) \quad (1)$$

where $T_{21}$ is the proportion of the mode power transferred between the two cores for propagation length $L$. Also, $\theta$ is the phase difference between the even and odd super-mode, defined by the relation

$$\theta = k_0 L \Delta n_{eff} \quad (2)$$

Here, $k_0$ is the wave number in free space and $\Delta n_{eff} = \left| n_{eff}(even) - n_{eff}(odd) \right|$ is the difference of the effective refractive indices for the two super-modes. Obviously, the interaction increases as the value of $\Delta n_{eff}$ increases too. In order to describe the strength of the interaction, the coupling length $L_C$ is introduced, defined by the relation

$$L_C = \frac{\lambda_0}{2\Delta n_{eff}} \quad (3)$$

where $\lambda_0$ is the free-space wavelength. Combining equations (1), (2) and (3) the proportion of the mode power, transferred between the cores, finally becomes

$$T_{21} = \sin^2\left(\frac{\pi}{2}\frac{L}{L_C}\right) \quad (4)$$

Consequently, the coupling length is equal to the propagation distance where the optical power is completely transferred from one core to the other. Obviously, the strength of the interaction is inversely proportional to the coupling length.

## 2. Analysis of the sensitivity of the MOF sensor

In order to use the MOF as refractive index sensors, the fluid under investigation fills the fiber holes altering the value of $\Delta n_{eff}$. This alteration can be practically measured either by the change of the transferred mode power for a monochromatic source, or by the shift of the wavelength when the maximum transmittance occurs for a broadband source [3, 18]. In the first case, the change of the transferred mode power as a function of the change in $\Delta n_{eff}$ can be expressed by the formula

$$\frac{\partial}{\partial n}T_{21} = \frac{\partial}{\partial n}\left[\sin^2\left(\frac{\pi}{2}\frac{L}{L_C}\right)\right]$$
$$= \sin\left(\pi\frac{L}{L_C}\right)\frac{\pi}{\lambda_0}L\frac{\partial}{\partial n}\Delta n_{eff} \quad (5)$$

where $n$ is the refractive index of the fluid under investigation. The quantity $\frac{\partial}{\partial n}T_{21}$ can be considered as a measure of the sensor sensitivity. Therefore, in a practical application $\frac{\partial}{\partial n}T_{21}$ should be maximized. This can be achieved under the condition that the fiber length becomes equal to $L_C/2$, or more generally $(2N+1)L_C/2$, where $N$ is a positive integer. In this case, Eq. (5) becomes

$$\frac{\partial}{\partial n}T_{21} = \pm\frac{\pi}{\lambda_0}L\frac{\partial}{\partial n}\Delta n_{eff}$$
$$= \pm\frac{\pi}{4}(2N+1)\frac{1}{\Delta n_{eff}}\frac{\partial}{\partial n}\Delta n_{eff} \quad (6)$$

In the second case, the fiber length is fixed to the value where maximum transmittance occurs, namely

$$L = (2N+1)L_C \quad (7)$$

Eq. (7) implies that the wavelength at maximum transmittance is equal to

$$\lambda_{max} = \frac{2\Delta n_{eff}L}{2N+1} \quad (8)$$

The shift of the wavelength for maximum transmittance as a function of the alteration $\Delta n_{eff}$ can be expressed through the formula

$$\frac{\partial}{\partial n}\lambda_{max} = \frac{2L}{2N+1}\frac{\partial}{\partial n}\Delta n_{eff}$$
$$= 2L_C\frac{\partial}{\partial n}\Delta n_{eff} \quad (9)$$
$$= \frac{\lambda_0}{\Delta n_{eff}}\frac{\partial}{\partial n}\Delta n_{eff}$$

The quantity $\frac{\partial}{\partial n}\lambda_{max}$ can also be regarded as a measure of the sensor sensitivity. Thus, in both cases the sensitivity of the sensor is proportional to the quantity

$$S = \frac{1}{\Delta n_{eff}}\frac{\partial}{\partial n}\Delta n_{eff} \quad (10)$$

## 3. Numerical calculation of the mode profile and the effective refractive index

In order to calculate the mode profile and the effective refractive index for each configuration, mode analysis has been applied on a cross section of the MOF [21]. In this analysis the MOF contains the two cores and the array of holes that confine the electromagnetic field. It is supposed that the electromagnetic wave propagates along the fiber length, defined as +z direction in the analysis. It has the form

$$E(x,y,z) = A(x,y)\exp(-\alpha_0 z)\exp(-i\beta z)$$
(11a)

$$H(x,y,z) = B(x,y)\exp(-\alpha_0 z)\exp(-i\beta z)$$
(11b)

where $E(x,y,z)$, $H(x,y,z)$ are the electric and magnetic field components of the propagating wave respectively, while the functions $A(x,y)$, $B(x,y)$ describe the mode profile. Here $\alpha_0$ is the linear loss coefficient and $\beta$ the propagation constant, related to the effective refractive index by the equation

$$\beta = k_0 n_{eff} \quad (12)$$

where $k_0 = 2\pi/\lambda_0$ is the free space wavenumber. Writing the Maxwell equations in matrix form, and eliminating the longitudinal field components, an eigenvalue problem is obtained. Its solution provides the propagation constants $\beta$ and the corresponding mode profiles for each configuration. More details on the algorithm used in the simulations can be found in [22]. The simulations were also assisted by the Comsol Multiphysics software package.

## 4. Improvement of the MOF sensor through the introduction of central holes at the fiber cores

Initially, a standard dual core micro-structured fiber has been considered. Three different models were studied with different values of the core separation, namely $2\Lambda$ - one hole between the cores, $3\Lambda$ - two holes between the cores, and $4\Lambda$ - three holes between the cores, where $\Lambda$ is the distance between adjacent holes. The results regarding the coupling length and the sensitivity are shown in figures 2a and 2b respectively. It is clear that as the wavelength increases, the coupling length decreases. This is expected since as the wavelength increases the mode profile is extended. Consequently, the interaction between the modes propagating in the two cores of the fiber increases too. Also, the coupling length increases as the core separation increases, which is also expected based on the above argument.

As far as the sensitivity is concerned it is also clear that it increases as the core separation becomes larger. It should be noted that the enhancement of the sensitivity as the core interaction becomes weaker, can be considered as a general trend, which is also valid for fibers with holes in the center of the cores.

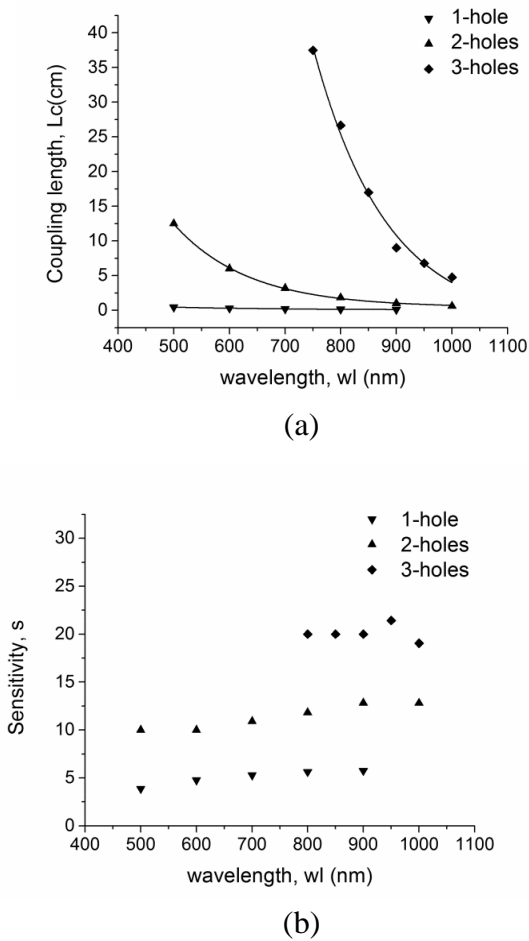

(a)

(b)

**Fig. 2:** The coupling length (a) and the sensitivity (b) as function of the wavelength for different values of the core separation. The solid lines are the fitting curves using single-exponential or linear decay functions.

As it was also mentioned in the introduction, it would be desirable to increase the dimensions of the fiber in order to facilitate the filling of the holes. In the numerical modeling this is achieved by introducing a parameter, scale, affecting all the dimensions of the fiber. For example, if the scale parameter is set to 2, then the dimensions of all the fiber features are doubled. The results have shown that the general trend observed in the case of scale=1, namely the increase of the sensitivity as the core separation increases, is also valid here. However, in this case the functional wavelength range of the fiber is restricted, either because the fiber becomes multi-mode, or the core interaction is fully suppressed. Of course, the functional wavelength range shrinks even further as the scale parameter increases, and eventually the fiber becomes unusable, at least as an RI sensor.

Therefore, it is clear that in order to use the MOF as a RI sensor for quite large values of the scale parameter, a substantial modification in the fiber design has to be made. This is necessary in order to enhance the core interaction and suppress the multimode behavior. Both of these goals can be achieved by introducing a small hole in the center of the fiber cores. This is described in the model by a parameter called ahole. For the precise, the parameter ahole describes the ratio of the hole diameter at the center of the cores to the hole diameter in the surrounding. From a physical point of view, the introduction of a hole at the center of the cores reduces the effective area of the modes in the core region. This leads to mode spreading, and consequently enhancement of the core interaction and suppression of the multimode behavior. This has also been verified by the results of the numerical simulations for 1-hole core separation, shown in Fig. 3a, 3b regarding the coupling length and the sensitivity respectively. Similar results are obtained for 2-hole core separation, as shown in Fig. 4.

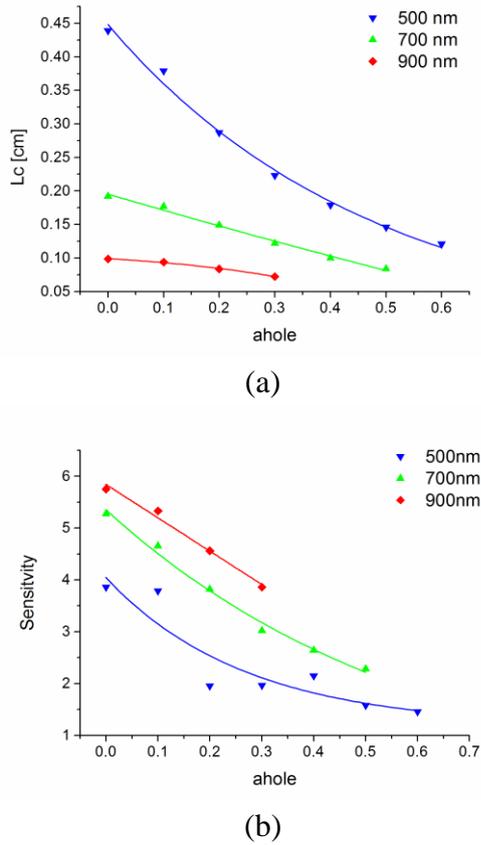

(a)

(b)

**Fig. 3**: The coupling length (a) and the sensitivity (b) as function of the central hole width for different wavelengths in the case of 1-hole separation between the fiber cores. The solid lines are the fitting curves using single-exponential or linear decay functions.

It is clear that the coupling length and the sensitivity decrease as the hole diameter increases, and consequently the mode interaction between the two cores becomes stronger, enabling the use of larger values of the scale parameter. As a characteristic example, the results for the sensitivity as a function of the parameter ahole, when the scale parameter is equal to 2 and 3 are shown in Fig. 5a and 5b respectively. It should be noted that in figures 2-5 there are not many data points for some wavelengths, due to the fact that the functional range of the sensor is limited at these wavelengths. However, in all cases, there are enough data to determine the form of the fitting functions.

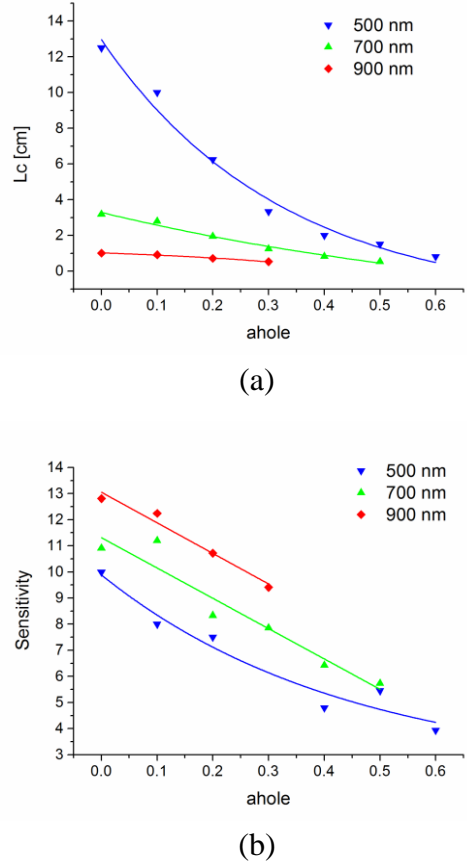

(a)

(b)

**Fig. 4**: The coupling length (a) and the sensitivity (b) as function of the central hole width for different wavelengths in the case of 2-hole separation between the fiber cores. The solid lines are the fitting curves using single-exponential or linear decay functions.

Further, numerical simulations have shown that, under certain conditions, the MOF can operates as an RI sensor for values of the scale parameter up to 10. Some characteristic values in the case of scale=5 for 1- and 2-hole separation between the fiber cores are shown in tables 1 and 2 respectively. From a practical point of view, this means that the micro-structured optical fiber can operates as an RI sensor for quite large values of the hole diameter, improving its usability.

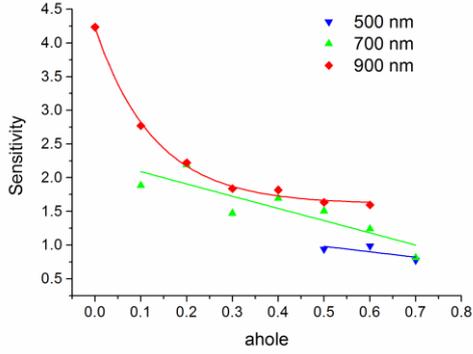

(a)

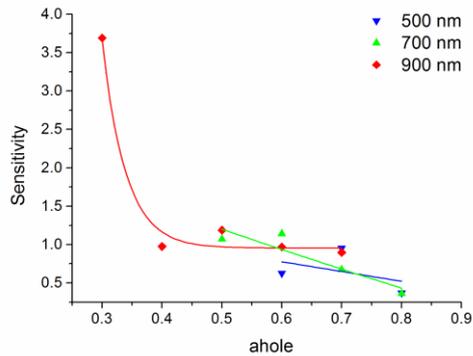

(b)

**Fig. 5**: The sensitivity as function of the central hole width for scale=2 (a) and scale=3 (b). The solid lines are the fitting curves using single-exponential or linear decay functions.

Further, numerical simulations have shown that, under certain conditions, the MOF can operates as an RI sensor for values of the scale parameter up to 10. Some characteristic values in the case of scale=5 for 1- and 2-hole separation between the fiber cores are shown in tables 1 and 2 respectively. From a practical point of view, this means that the micro-structured optical fiber can operates as an RI sensor for quite large values of the hole diameter, improving its usability. Further, the high values of the sensitivity obtained, indicate that the system can detect very subtle changes in the refractive index of its environment.

**Table 1:** The functional range of a MOF sensor in the case of scale=5 and 1-hole core-separation.

| ahole | λ[nm] | S | Lc[m] |
|---|---|---|---|
| 0.85 | 500 | 12.6 | 0.026 |
| 0.7 | 600 | 0.91 | 0.027 |
| 0.7 | 700 | 0.67 | 0.023 |
| 0.8 | 700 | 0.27 | 0.018 |
| 0.7 | 800 | 0.98 | 0.019 |
| 0.8 | 800 | 1.63 | 0.016 |
| 0.6 | 900 | 0.50 | 0.022 |

**Table 2:** The functional range of a MOF sensor in the case of scale=5 and 2-hole core separation.

| ahole | λ[nm] | S | Lc[m] |
|---|---|---|---|
| 0.9 | 500 | 12.5 | 0.063 |
| 0.7 | 700 | 10.0 | 0.175 |
| 0.8 | 800 | 3.64 | 0.073 |
| 0.8 | 900 | 1.43 | 0.064 |

For example, in the case of 2-hole core separation, scale=5, ahole=0.7 and $\lambda = 700nm$, it has been found that $S = 10$. Thus, according to Eq. (9) the wavelength shift is $\frac{\partial}{\partial n}\lambda_{max} = 7000\frac{nm}{RIU}$, which an impressive result is. From a practical point of view, given that a high-end optical spectrum analyzer can easily detect wavelength changes of the order of 0.05 nm, the above result indicates that the minimum refractive index change detectable by the system is $7 \times 10^{-6}$. Thus, the system can detect even the subtlest changes in its environment, making the sensor suitable for very demanding applications, including biological ones, as it will be discussed in the next section.

## 5. Application of the system as bio-sensor

In recent years, the use of nano-photonic systems, and especially photonic crystal fibers as biosensors, has become a subject of intense research [22-28]. In this section, we shall check the potential of our system as a bio-sensor by considering a thin layer of biomaterial attached in the inner surface of the core holes. For example, a practical use of this analysis could be the study and optimization of label-free antibody detection using the highly selective antigen-antibody binding [29]. Specifically, we have studied the effect of the bio-layer thickness $t_b$ and refractive index $n_b$ on $\Delta n_{eff}$. Clearly, the sensitivity of the system as a biosensor is proportional to the quantities

$$S_{tb} = \frac{1}{\Delta n_{eff}} \frac{\partial \Delta n_{eff}}{\partial t_b}\bigg|_{nb} \quad (13a)$$

as far as the layer thickness is concerned and

$$S_{nb} = \frac{1}{\Delta n_{eff}} \frac{\partial \Delta n_{eff}}{\partial n_b}\bigg|_{tb} \quad (13b)$$

as far as the refractive index of the bio-layer is concerned. Through extensive numerical simulations we have found that the slopes $\frac{\partial \Delta n_{eff}}{\partial t_b}\bigg|_{nb}$ and $\frac{\partial \Delta n_{eff}}{\partial n_b}\bigg|_{tb}$ are constant over quite a wide range of the layer thickness (0-100 nm) and refractive index (1.35-1.45). Some characteristic plots are shown in Fig. 6.

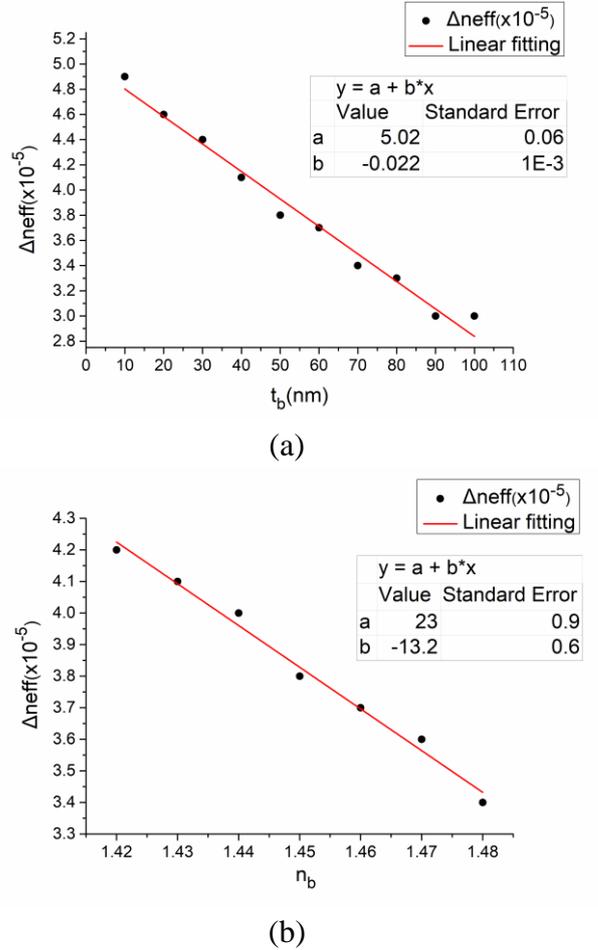

(a)

(b)

**Fig. 6:** The dependence of $\Delta n_{eff}$ on (a) the thickness and (b) the refractive index of the bio-layer. The graphs correspond to the case of 2-hole separation between the cores, ahole=0.55, while Fig. 6a corresponds to $n_b$=1.45 and Fig. 6b to $t_b$=50 nm. Further, in both cases it is assumed that the biolayer is developed only on the central holes of the cores while the rest of the holes is filled with water. All simulations have been performed for a wavelength of 633 nm, which corresponds to minimum losses for PMMA fibers.

It is clear that $\Delta n_{eff}$ decreases linearly as the layer thickness or refractive index increases, which is expected because the propagating mode is shielded inside the core holes decreasing the interaction between the fiber cores.

From the numerical values of the slopes we can calculate the sensitivity of the sensor which, according to Eq. (9), is 3.5 nm/nm when the sensor is used to detect thickness changes (fig. 6a) and $2.2 \times 10^3 \, nm/RIU$ when the sensor is used to detect refractive index changes (fig. 6b). Both these results correspond to a bio-layer of 50 nm thickness and 1.45 refractive index. They can improve even further as the layer thickness and/or refractive index increase, as it will be shown later.

Given that a modern high-end optical spectrum analyzer can easily detect wavelength shifts of the order of 0.05 nm, it can easily be deduced that the bio-sensor can detect thickness changes of the order of $1.4 \times 10^{-2} \, nm$ and refractive index changes of the order of $2.3 \times 10^{-5}$. These results are impressive, indicating that the system can detect even the subtlest changes regarding the composition (through changes in the refractive index) and/or the thickness of the bio-layer.

Further, it has been found that the dependence of the slopes $\left.\frac{\partial \Delta n_{eff}}{\partial t_b}\right|_{nb}$ and $\left.\frac{\partial \Delta n_{eff}}{\partial n_b}\right|_{tb}$ on the refractive index and layer thickness respectively can be described by simple linear or single exponential functions. Some characteristic plots are shown in Fig. 7. These results suggest that the calibration of the system for use as a biosensor should be especially easy. It is also clear that the sensitivity of the system increases as the layer thickness and/or refractive index increase, giving the potential to improve even further the impressive results regarding the detection limit of the system. Finally, it should be noted that these trends are valid over a wide range of the sensor parameters, namely core separation, central hole diameter, and scaling. Some characteristic plots for a different set of parameters are shown in Fig. 8. It is obvious that in this case the dependence of $\left.\frac{\partial \Delta n_{eff}}{\partial t_b}\right|_{nb}$ on the refractive index of the biolayer is not linear, but it is described by a single exponential growth function. However, the behavior of the system can still be easily modeled and predicted.

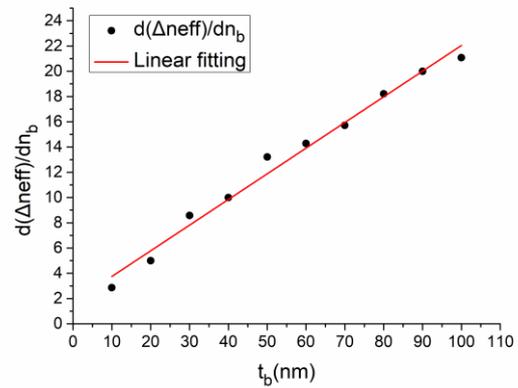

(a)

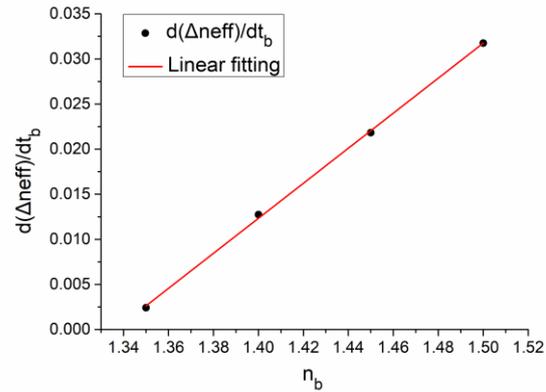

(b)

**Fig. 7:** The dependence of the slopes (a) $\left.\frac{\partial \Delta n_{eff}}{\partial t_b}\right|_{nb}$ and (b) $\left.\frac{\partial \Delta n_{eff}}{\partial n_b}\right|_{tb}$ on the refractive index and layer thickness respectively. The parameters used in the simulations are the same as in Fig. 6.

Thus, the design parameters of the sensor can be tailored to the demands of specific applications, without losing its main advantages, namely high sensitivity and ease of calibration.

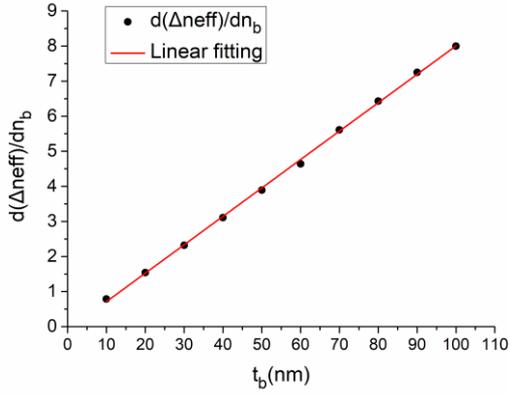

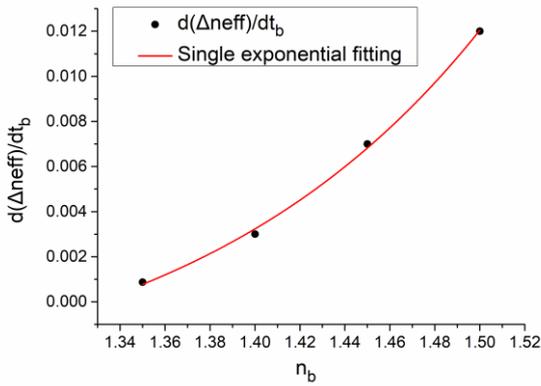

**Fig. 8:** The dependence of the slopes (a) $\left.\frac{\partial \Delta n_{eff}}{\partial t_b}\right|_{nb}$ and (b) $\left.\frac{\partial \Delta n_{eff}}{\partial n_b}\right|_{tb}$ on the refractive index and layer thickness respectively. The parameters used in the simulations are the same as in Fig. 7 except the core separation, which is now 1-hole.

## 6. Conclusions

In conclusion, a new design for refractive index sensors based on a micro-structured dual core optical fiber is exhibited. This is based on the introduction of a small hole at the central region of the fiber cores. This causes mode spreading, enhancing the core interaction and suppressing undesirable multi-mode behavior. In this way, the dimensions of the MOF sensor can become large, facilitating its practical use. Further, the same principle has been applied to a dual core conventional optical fiber. In this case, numerical simulations have shown that it can also operate as a refractive index sensor, where the sensitivity can be controlled mainly by the core separation. Further, we have performed simulations regarding the potential of the system for use as a bio-sensor, which gave very promising results regarding both the sensitivity and the ease of the calibration.

## Acknowledgment

We would like to thank Prof. A. Boudouvis for providing access to the Comsol Multiphysics software package.

## References

[1] C. M. B. Cordeiro, M. A. R. Franco, G. Chesini, E. C. S. Barretto, R. Lwin, C. H. B. Cruz1, and M. C. J. Large , "Microstructured-core optical fibre for evanescent sensing applications", Opt. Express 14 (2006) 13056

[2] D. Wu, B. Kuhlmey, and B. Eggleton, "Ultrasensitive photonic crystal fiber refractive index sensor," Opt. Lett. 34 (2009) 322

[3] W. Yuan, G. E. Town, and O. Bang, "Refractive index sensing in an all-solid twin-core photonic bandgap fiber", IEEE Sens. J. 10 (2010) 1192

[4] G. Town, W. Yuan, R. McCosker, and O. Bang, "Microstructured optical fiber refractive index sensor," Opt. Lett. 35 (2010) 856

[5] H. W. Lee, M. A. Schmidt, P. Uebel, H. Tyagi, N. Y. Joly, M. Scharrer and P. St.J. Russell, "Optofluidic refractive-index sensor in step-index fiber with parallel hollow micro-channel", Opt. Express 19 (2011) 8200

[6] B. Sun, M.-Y. Chen, Y.-K. Zhang, J. Yang,1 J. Yao, and H.-X. Cui, "Microstructured-core photonic-crystal fiber for ultra-sensitive refractive index sensing", Opt. Express 19 (2011) 4091

[7] A. P. Zhang, G. Yan, S. Gao, S. He, B. Kim, J. Im, and Y. Chung, "Microfluidic refractive-index sensors based on small-hole microstructured optical fiber Bragg gratings", Appl. Phys. Lett. 98 (2011) 221109

[8] M. Frosz, A. Stefani, and O. Bang, "Highly sensitive and simple method for refractive index sensing of liquids in microstructured optical fibers using four-wave mixing," Opt. Express 19 (2011) 10471

[9] S. Warren-Smith and T. Monro, "Exposed core microstructured optical fiber Bragg gratings: refractive index sensing," Opt. Express 22 (2014) 1480

[10] P. Torres, E. Reyes-Vera, A. Díez, and M. Andrés, "Two-core transversally chirped microstructured optical fiber refractive index sensor," Opt. Lett. 39 (2014) 1593

[11] Z. Li, C. Liao, Y. Wang, X. Dong, S. Liu, K. Yang, Q. Wang, and J. Zhou, "Ultrasensitive refractive index sensor based on a Mach-Zehnder interferometer created in twin-core fiber," Opt. Lett. 39 (2014) 4982

[12] W. Dong, J. Wei, X. Wang, Z. Kang, and X. Xu, "Liquid refractive index sensor based on polymer fiber with micro-holes created by femtosecond laser," Chin. Opt. Lett. 12 (2014) 090601

[13] L. Rindorf, J. B. Jensen, M. Dufva, L. H. Pedersen and P. E. Høiby, "Photonic crystal fiber long-period gratings for biochemical sensing", Opt. Express 14 (2006) 8224

[14] D. Passaro, M. Foroni, F. Poli, A. Cucinotta, S. Selleri, J. Lægsgaard, and A. O. Bjarklev, "All-silica hollow-core microstructured Bragg fibers for biosensor application", IEEE Sens. J. 8 (2008) 1280

[15] J. R. Ott, M. Heuck, C. Agger, P. D. Rasmussen, and O. Bang, "Label-free and selective nonlinear fiber-optical biosensing", Opt. Express 16 (2008) 20834

[16] K. Milenkoa, D.J.J. Hub, P.P. Shumc and T.R. Wolinskia, "Hollow-core Bragg fiber for bio-sensing applications", Acta Phys. Pol. A 118 (2010) 1205

[17] E. Coscelli, M. Sozzi, F. Poli, D. Passaro, A. Cucinotta, S. Selleri, R. Corradini, and R. Marchelli, "Toward A Highly Specific DNA Biosensor: PNA-Modified Suspended-Core Photonic Crystal Fibers", IEEE J. Sel. Top. Quant. 16 (2010) 967

[18] C. Markos, W. Yuan, K. Vlachos, G. E. Town, and O. Bang, "Label-free biosensing with high sensitivity in dual-core microstructured polymer optical fibers", Opt. Express 19 (2011) 7790

[19] Y.S. Skibina, V.V. Tuchin, V.I. Beloglazov, G. Steinmeyer, J. Bethge, R. Wedell, N. Langhoff, "Photonic crystal fibres in biomedical investigations", Quantum Electron. 4 (2011) 284

[20] T. Yadav, R. Narayanaswamy, M. Abu Bakar, Y. Kamil, and M. Mahdi, "Single mode tapered fiber-optic interferometer based refractive index sensor and its application to protein sensing," Opt. Express 22 (2014) 22802

[21] J. M. Jin, "The Finite Element Method in Electromagnetics", 3$^{rd}$ Edition, Wiley-IEEE Press, 2014

[22] G. N. Tsigaridas, "A study on refractive index sensors based on optical micro-ring resonators", Photonic Sens 7 (2017) 217

[23] M.F.H. Arif, K. Ahmed, S. Asaduzzaman and M.A.K. Azad, "Design and optimization of photonic crystal fiber for liquid sensing applications", Photonic Sens (2016) 6: 279. Photonic Sens 6 (2016) 279

[24] M. Nejadebrahimy, L. Halimi and H.

Alipour-Banaei, "Design and simulation of ultrasensitive nano-biosensor based on OFPC", Photonic Sens 5 (2015) 43

[25] S. Olyaee, S. Najafgholinezhad, and H. Alipour Banaei, "Four-channel label-free photonic crystal biosensor using nanocavity resonators", Photonic Sens 3 (2013) 231

[26] M. Consales, M. Pisco and A. Cusano, "Lab-on-fiber technology: a new avenue for optical nanosensors", Photonic Sens 2 (2012) 289

[27] O. Frazão, R. M. Silva, M. S. Ferreira et al., "Suspended-core fibers for sensing applications", Photonic Sens 2 (2012) 118

[28] L. Zhang, J. Lou and L. Tong, "Micro/nanofiber optical sensors", Photonic Sens 1 (2011) 31

[29] J. B. Jensen, P. E. Hoiby, G. Emiliyanov, O. Bang, L. Pedersen, and A. Bjarklev, "Selective detection of antibodies in microstructured polymer optical fibers", Opt. Express 13 (2005) 5883